\documentclass[prc,aps,twocolumn,nofootinbib,showpacs,showkeys]{revtex4}
\usepackage{graphicx,epsf,amsfonts,amssymb,amsbsy}
\newcommand{\be} {\begin{eqnarray*}}  
\newcommand{\ee} {\end{eqnarray*}}  
\newcommand{\bi} {\begin{itemize}}  
\newcommand{\ei} {\end{itemize}}  
  
\newcommand{\bcen}{\begin{center}}  
\newcommand{\ecen}{\end{center}}  
\newcommand{\beq}{\begin{equation}}  
\newcommand{\eeq}{\end{equation}}  
\newcommand{\bea}{\begin{eqnarray}}  
\newcommand{\eea}{\end{eqnarray}}  
\newcommand{\ba}{\begin{array}}  
\newcommand{\ea}{\end{array}}  
\newcommand{\bann}{\begin{eqnarray*}}  
\newcommand{\eann}{\end{eqnarray*}}

\begin{document}  

\title{Color-spin locking phase in two-flavor quark matter  
for compact star phenomenology}  
  
\author{D. N. Aguilera}
\email{deborah.aguilera@uni-rostock.de}  
\affiliation{Institut f\"ur Physik, Universit\"at Rostock, \\  
Universit\"atsplatz 3, 18051 Rostock, Germany}
\author{D. Blaschke}
\email{blaschke@gsi.de}
\affiliation{Gesellschaft f\"ur Schwerionenforschung (GSI), \\  
Planckstr. 1, 64291 Darmstadt, Germany\\
Bogoliubov  Laboratory of Theoretical Physics, JINR Dubna,\\  
Joliot-Curie Street 6, 141980  Dubna, Russia}   
\author{M. Buballa}
\email{michael.buballa@physik.tu-darmstadt.de}
\affiliation{Institut f\"ur Kernphysik, TU Darmstadt,\\  
Schlossgartenstr. 9, 64289 Darmstadt, Germany}    
\author{V.L. Yudichev} 
\email{yudichev@thsun1.jinr.ru}
\affiliation{Bogoliubov  Laboratory of Theoretical Physics, JINR Dubna,\\  
Joliot-Curie Street 6, 141980  Dubna, Russia}   
\begin{abstract}  
We study a spin-1 single flavor color superconducting phase which 
results from a color-spin locking (CSL) interaction in two-flavor 
quark matter.  
This phase is particularly interesting for compact star cooling 
applications since the CSL phase may survive under charge neutrality  
constraints implying a mismatch between up- and down-quark chemical 
potentials which can destroy the scalar diquark condensate.  
CSL gaps are evaluated within an NJL model and they are found to be consistent with cooling phenomenology if a density dependent coupling constant is used.
\end{abstract}  

\pacs{12.38.Mh, 24.85.+p,26.60.+c, 97.60.Jd}  
\keywords{Color superconductivity, spin-one condensates, compact star cooling}  
\maketitle  
\section{Introduction}  
  
Theoretical investigations of the QCD phase diagram at low temperatures  
$T\ll T_c$ and high baryochemical potential $\mu_B\sim 1$ GeV have recently  
received new impulses when color superconducting quark matter phases with  
large diquark pairing gaps of about 100 MeV were suggested within  
nonperturbative low-energy QCD models 
\cite{Rapp:1997zu,Alford:1997zt,Blaschke:1998md}.  
Prior to these works, color superconductivity in quark matter has been 
discussed using perturbative one-gluon exchange interactions resulting 
in small diquark pairing gaps of about 1 MeV only  
\cite{barrois77,bailin84}.  
More systematic studies of the possible diquark pairing patterns have revealed  
a very rich phase structure of cold quark matter, see e.g.  
\cite{Alford:2001dt,Buballa:2003qv,Schmitt:2004et,Ruster:2005jc,Blaschke:2005uj}.  
The most prominent color superconducting phases with large diquark 
pairing gaps are the two-flavor scalar diquark condensate (2SC) and 
the color-flavor locking (CFL) condensate. The latter requires 
approximate SU(3) flavor symmetry and occurs therefore only at rather 
large quark chemical potentials, $\mu_q \gtrsim 410-500$ MeV  
\cite{Ruster:2005jc,Blaschke:2005uj,Neumann:2002jm,Oertel:2002pj}. 
 
The most favorable places in nature where color superconducting states of 
matter are expected to occur are the interiors of compact stars, with 
temperatures well below 1 MeV and central densities exceeding the 
nuclear saturation density by about one order of magnitude. 
However, a closer examination of the compact star constraints on phases of 
dense matter reveals that the requirement of charge neutrality 
\cite{Alford:2002kj}  may inhibit flavor asymmetric pair condensation 
in the 2SC phase and even influence the CFL phase border. 
As the quark chemical potential corresponding to the highest possible 
central densities may barely reach the above constraint for the 
occurrence of strange quark matter \cite{Gocke:2001ri}, it is expected 
that the volume fraction of a strange quark matter phase in a compact 
star cannot be sufficient to entail observable consequences.  
The two-flavor quark matter phases, on the other hand, may occupy a large 
volume \cite{Grigorian:2003vi,Shovkovy:2003ce}, but due to the 
flavor asymmetry the occurrence of the 2SC phase is rather model dependent  
\cite{Ruster:2005jc,Blaschke:2005uj,Aguilera:2004ag}.  
 
Most sensitive observable consequences of quark matter phases in compact stars 
come from the cooling evolution and it has been demonstrated that the 
occurrence of unpaired quarks in the core of hybrid stars leads to 
rapid cooling via the direct Urca process  
\cite{Blaschke:1999qx,Page:2000wt,Blaschke:2000dy}. 
To describe existing cooling data for isolated compact stars the 
occurrence of quark pairing in a 2SC phase is not sufficient and a 
residual weak pairing channel with a small gap $\Delta_X\sim 10 - 100$ 
keV has been introduced phenomenologically \cite{Grigorian:2004jq}. 
No microscopic calculation for the justification of this 2SC+X phase 
could be presented up to now.  
 
Given this fact and that the 2SC phase is very fragile with respect to 
the flavor asymmetry, the question arises whether 
there exist other pairing patterns in two-flavor quark matter which  
survive compact star constraints and fulfill the requirements from the 
cooling phenomenology that all quarks shall be paired with the 
smallest pairing gap being of the order of several keV. 
Single flavor spin-1 pairs are good candidates since they are inert 
against large splittings in the quark Fermi levels for different 
flavors 
occurring due to electroneutrality 
They  have been introduced first in Refs.  
\cite{Schafer:2000tw,Alford:2002rz} and their properties have been  
investigated later more in detail, see  \cite{Schmitt:2004et} and 
Refs. therein. 
Recently, a specific spin-1 pairing pattern -- the so-called
A phase -- has been employed to suggest a new mechanism for the 
origin of pulsar kicks \cite{Schmitt:2005ee}.
The idea is based on the fact that the pairing in the A phase is
anisotropic, giving rise to a directed neutrino emission during the
cooling process.
On the other hand, there are still ungapped modes in the A phase,
which would again lead to difficulties with cooling phenomenology.

Therefore, in the present work we want to argue in favor of 
spin-1 pairing in the color spin locked (CSL) phase as a possible 
candidate. This phase has recently been found to be the most stable
spin-1 phase, at least at very high densities~\cite{Schmitt:2004et}.
As we will show, as long as the quarks do not become exactly massless,
all modes are gapped in the CSL phase.
To study this more quantitatively, we present a microscopic calculation 
within an NJL-type model, thus providing results for future phenomenological 
applications, like compact star cooling \cite{Grigorian:2004jq}.

  
\section{NJL model for the CSL phase}  

The CSL pairing pattern we want to study is defined by the equality 
of three diquark condensates,
\beq  
\langle q_f^{T}~C\gamma^3\lambda_2~q_f \rangle  
= \langle q_f^{T}~C\gamma^1\lambda_7~q_f \rangle  
= \langle q_f^{T}~C\gamma^2\lambda_5~q_f \rangle \equiv \eta_f~,
\label{csl}
\eeq 
where $q_f$ is a quark field of flavor $f$ with three color and, of course,
two spin degrees of freedom. The operators $\lambda_A$ are antisymmetric
Gell-Mann matrices acting in color space, while $\gamma^\mu$ are Dirac 
matrices, and $C = i\gamma^2\gamma^0$ is the matrix of charge conjugation.
Hence, each of the three diquark condensates belongs to the antisymmetric
antitriplet in color space and, at the same time, represents a vector
component in spin-space. This means, both, color $SU(3)$ and rotational 
$O(3)$, are broken for $\eta_f \neq 0$. There remains, however, a residual 
invariance under a common transformation in which color and spin rotations
are locked to each other.

Together with the diquark condensates, we consider the 
presence of the quark-antiquark condensates
\beq
     \sigma_f = \langle \bar q_f q_f \rangle~,  
\label{sigmaf}
\eeq
which are responsible for dynamical chiral symmetry breaking in vacuum.
In the 
density regime we are interested in, these condensates are
relatively small. However, as we will see, it is important to keep them
in the calculations.

In order to investigate the properties of this condensation pattern,
we consider a two-flavor system of quarks, $q = (u,d)^T$, 
and employ an NJL-type Lagrangian
\beq 
\mathcal{L}_{{\rm eff}}= \mathcal{L}_0 + \mathcal{L}_{q\bar q}
+\mathcal{L}_{qq}~, 
\eeq 
with a free part
\beq 
\mathcal{L}_0=\bar q\left(i\not\!\partial-m \right)q~, 
\eeq
and 4-point interactions
in the channels corresponding to Eqs.~(\ref{csl}) and (\ref{sigmaf}):
\bea
\mathcal{L}_{q\bar q}&=& G \sum_{a=0}^3(\bar q\,\tau_a\,q)^2 \,+\, \dots~,
\label{barqq}\\ 
\mathcal{L}_{q q}&=& 
- H_v 
\left(\bar q\,\gamma^{\mu} C\,\tau_{\rm S} \lambda_{\rm A}\,\bar q^T \right) 
\left(q^T C \gamma_{\mu}\,\tau_{\rm S} \lambda_{\rm A}\,q \right) \,+\, 
\dots~ 
\label{qq} 
\eea 
Here the matrices $\tau_a$, $a = 1,2,3$, are the usual
Pauli matrices in flavor space while $\tau_0$ is a unit matrix.  
As before, Gell-Mann matrices in color space are symbolized by $\lambda$.
The subscripts A and S denote antisymmetric and symmetric matrices, respectively,
i.e., $\tau_{\rm S} \in \{\tau_0,\tau_1,\tau_3\}$ and 
$\lambda_{\rm A} \in \{\lambda_2,\lambda_5,\lambda_7\}$.
As usual, repeated indices are summed over.

$\mathcal{L}_{q\bar q}$ and $\mathcal{L}_{qq}$
should be interpreted as effective interactions
to be used in mean-field (Hartree) approximation. 
In general, there could be further interaction terms,
indicated by the ellipsis in Eqs.~(\ref{barqq}) and (\ref{qq}).
For instance, $\mathcal{L}_{qq}$ could contain a scalar term in the
flavor singlet color antitriplet channel, corresponding to 
standard 2SC pairing. 
However, as outlined in the Introduction, we assume that under 
compact star conditions the mismatch between up and down quark Fermi momenta
is too large to allow for cross-flavor condensation.
In this case, the corresponding interaction terms do not contribute
to the mean-field thermodynamics. 
In the same way, we assume that there is no significant condensation
in any other channel except $\eta_f$ and $\sigma_f$.

As a specific example, we could think of $\mathcal{L}_{q\bar q}$ and 
$\mathcal{L}_{qq}$ to arise via a Fierz transformation from a color-current 
interaction
$\mathcal{L}_{int} = -g \left(\bar q\,\gamma^{\mu} \lambda_a\,q\right) 
                        \left(\bar q\,\gamma_{\mu} \lambda_a\,q\right)$,
corresponding to the quantum numbers of a single gluon exchange.
In this case the ratio of the two coupling constants is given by
$G:H_v=1:\frac38$. 
Later, we will employ this relation in order to constrain the number of 
parameters in the numerical calculations. 
At the present stage, however, we can leave $G$ and $H_v$ as arbitrary
constants. 

In order to derive the mean-field thermodynamic potential for an
ensemble of up and down quarks at temperature $T$ and chemical
potentials $\mu_u$ and $\mu_d$, respectively,
we linearize $\mathcal{L}_{{\rm eff}}$ in the 
presence of the condensates $\eta_f$ and $\sigma_f$.
One finds that the different flavors decouple, i.e., the
thermodynamic potential is given by the sum
\beq  
\Omega_q(T,\{\mu_f\}) =  \Omega_u(T,\mu_u) +  \Omega_d(T,\mu_d)~.  
\label{omegaq} 
\eeq
Defining the constituent quark masses and the diquark gaps  
\beq 
M_f = m_f-4G\sigma_f~, \quad 
\Delta_f = 4 H_v \eta_f~, 
\label{Delta_f} 
\eeq 
and using the Nambu-Gorkov bispinors 
\beq 
\begin{array}{cc} 
\psi=\frac{1}{\sqrt{2}}\left(  
\begin{array}{c}  
q\\ 
C\,\bar q^T 
\end{array}  
\right),~~  
& 
~~\bar \psi=\frac{1}{\sqrt{2}}\left(  
\begin{array}{cc}  
\bar q,& 
q^T C 
\end{array}  
\right)  
\end{array} 
\eeq   
the contributions for each flavor can be written as  
\bea  
\Omega_f(T,\mu_f)&=& 
-T \sum_{n} \int \frac{d^3p}{(2\pi)^3}  
\frac{1}{2}{\rm Tr}\ln({\frac{1}{T}S^{-1}_f(i\omega_n,\vec p\,)})
\nonumber\\
&+& 
\frac{1}{8G} (M_f-m)^2
+ \frac{3}{8H_v} |\Delta_f|^2
~,  
\label{omegaf}
\eea  
where the sum is over fermionic Matsubara frequencies 
$\omega_n = (2n+1)\pi T$. 
The inverse of fermion propagator 
is given by  
\beq  
S^{-1}_f(p)= 
\left(  
\begin{array}{cc}  
 \not\!p +\mu_f\gamma^0-M_f&  
\hat\Delta_f\\
-{\hat\Delta_f}^\dagger&
\not\!p -\mu_f\gamma^0-M_f  
\end{array}  
\right)
\eeq
in Nambu-Gorkov space, where $\hat\Delta_f=\Delta_f(\gamma^3\lambda_2+\gamma^1\lambda_7+\gamma^2\lambda_5)$. Taking into account its color and Dirac structure,
this is a $24 \times 24$  matrix. 
Explicit evaluation of the trace yields
\beq
{\rm Tr}\ln\!\left(\!\frac{S^{-1}_f(p)}{T}\!\right)
= 2 \left(\ln\frac{F_f^-(p)}{T^2} + \ln\frac{F_f^+(p)}{T^2} +
          \ln\frac{G_f(p)}{T^8}\right)\!,
\eeq
where 
\beq 
F_f^\mp(p)
= p^2_0- \left(
\varepsilon_f^2 + \mu_f^2 + |\Delta_f|^2 \mp 2\sqrt{\mu_f^2\varepsilon_f^2
+|\Delta_f|^2{\vec p\,}^{2}} \right)
\label{F_function} 
\eeq
and 
\bea 
G_f(p)&=&\left(p_0^2-\left(\varepsilon_f-\mu_f\right)^2\right)^2 
\left(p_0^2-\left(\varepsilon_f+\mu_f\right)^2\right)^2\nonumber\\ 
&-&10|\Delta_f|^2 A 
\left(p_0^2-\left(\varepsilon_f-\mu_f\right)^2\right) 
\left(p_0^2-\left(\varepsilon_f+\mu_f\right)^2\right) 
\nonumber\\ 
&-&33|\Delta_f|^4 B_+B_--40|\Delta_f|^6 A+16|\Delta_f|^8~. 
\label{G_function} 
\eea 
Here, we have introduced the abbreviations
$\varepsilon_f^2={\vec p\,}^{2}+M_f^2$ and 
\bea 
A&=&p_0^2-\frac{3}{5}\,{\vec p\,}^{2}-\mu_f^2-M_f^2~,\\ 
B_{\pm}&=&p_0^2-\frac{25}{33}\,{\vec p\,}^{2}-\mu_f^2-M_f^2\nonumber\\
&\pm& \frac{2}{33}\sqrt{16\,{\vec p\,}^{4}+297\,\mu_f^2{\vec p\,}^{2}+561\,M_f^2\mu_f^2}~. 
\eea 
Finally, we have to turn out the Matsubara sum. 
To that end we determine the zeros $E^2_{f;k}$ of the polynomials $F_f^\mp$ 
and $G_f$ with respect to $p_0^2$,
\bea 
F_f^\mp(p_0,\vec p\,)&=& p_0^2-E^2_{f;1,2}~, 
\label{FF} 
\\ 
G_f(p_0,\vec p\,)&=& \prod_{k=3}^{6}(p_0^2-E^2_{f;k})~, 
\label{GG} 
\eea  
and apply the formula  
\bea 
T \sum_n \ln\left(\frac{1}{T^2}(\omega_n^2+\lambda_k^2) \right)=\lambda_k 
+2T\ln\left(1+e^{-\lambda_k/T}\right)~ 
\eea 
to get the result
\bea  
\Omega_f(T,\mu_f)&=&  
\frac{1}{8G} (M_f-m)^2
+ \frac{3}{8H_v} |\Delta_f|^2
\nonumber\\ 
&-&
\sum_{k=1}^6 \int \frac{d^3p}{(2\pi)^3}  
\left(E_{f;k}+2T\ln{(1+e^{-E_{f;k}/T})}\right).\nonumber\\ 
\label{omega} 
\eea  
This expression is the basis for the further analysis which will be
performed on the mean-field level, i.e.  at the
stationary points 
\beq
    \frac{\delta\Omega_f}{\delta\Delta_f} = 0~, \quad
    \frac{\delta\Omega_f}{\delta M_f} = 0~,
\eeq
defining a set of gap equations for $\Delta_f$ and $M_f$. 
Among the solutions, the stable one is the solution
which corresponds to the absolute minimum of $\Omega_f$.

  
\section{Dispersion laws}  
\label{dispers}

The six functions $E_{f,k}$, defined in Eqs.~(\ref{FF}) and (\ref{GG})
correspond to the dispersion laws of the quasiparticle states of flavor
$f$. 
In the above discussions we have assumed that we know these functions.
This is obviously the case for $E_{f;1}$ and $E_{f;2}$ which are directly
read off from Eq.~(\ref{F_function}):
\beq 
E_{f;1,2}
=\sqrt{
\varepsilon_f^2 + \mu_f^2 + |\Delta_f|^2 \mp 2\sqrt{\mu_f^2\varepsilon_f^2
+|\Delta_f|^2{\vec p\,}^{2}}}~. 
\label{e12}
\eeq 
Here we have chosen the positive square-root, corresponding to the 
quasiparticle energy, while the negative square-root corresponds to a 
quasihole state.

It is advantageous to define effective quantities,
\bea 
\varepsilon_{f, {\rm eff}}^2 &=& \vec p\,^2+M_{f,{\rm eff}}^2~,\\
M_{f,{\rm eff}} &=& \frac{\mu_f}{\mu_{f,{\rm eff}}}M_f~, \\
\mu_{f,{\rm eff}}^2 &=& \mu_f^2+|\Delta_f|^2~,
\eea  
and
\beq
\Delta_{f,\rm eff} 
= \frac{M_f}{\mu_{f,{\rm eff}}}\,|\Delta_f|~ 
\label{Deltaeff}
\eeq 
in order to bring Eq.~(\ref{e12}) into the standard form
\bea  
E_{f;{1,2}}&=& \sqrt{(\varepsilon_{f,{\rm eff}}\mp\mu_{f,{\rm eff}})^2+  
\Delta_{f,{\rm eff}}^2}~. 
\label{E_1,2} 
\eea 
Note that, in contrast to the gap parameter $\Delta_f$, 
the effective gap $\Delta_{f,\rm eff}$ corresponds to the true gap
in the spectrum of the first quasiparticle mode. 
In fact, as we will see below, it is the smallest gap of all
quasiparticle modes in our model. It is therefore the relevant quantity 
for cooling and other transport properties.

Compared with $\Delta_f$, the effective gap $\Delta_{f,\rm eff}$ is 
suppressed by an additional factor of the order $M_f/\mu_f$. 
This implies that the mode $E_{f,1}$ becomes gapless in the chiral limit    
when $M_f =0$. This is the reason why the current quark mass $m$ and 
the chiral condensate $\sigma_f$ should be retained,
even if they are small.  

The four remaining dispersion relations $E_{f;k},~k=3 \dots 6$ are the zeros 
of the fourth-order polynomial $G_f(p)$ (\ref{G_function}).
Although, in principle, the exact solutions can be determined analytically,
the resulting expressions are extremely long and difficult to handle.
In practice, it is therefore more convenient to perform a numerical
search. 

\begin{figure}[bth]  
  \begin{center}  
    \includegraphics[width=0.82\linewidth,angle = -90]{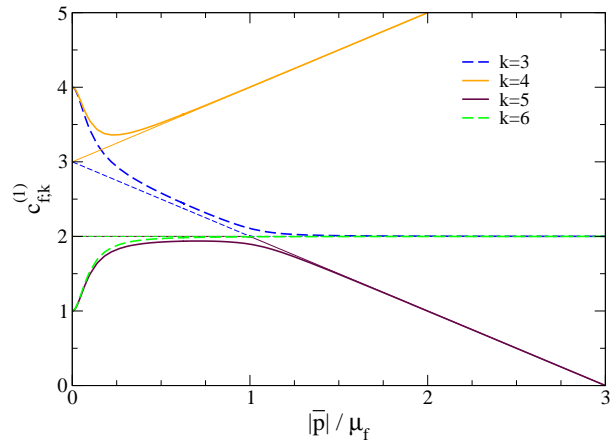}  
    \caption{Dispersion relation coefficients $c^{(1)}_{f;k}$, $k=3...6$, 
 as functions of $|\vec p|/\mu_f$ for fixed $\mu_f=400$ MeV.  
The thick lines correspond to a typical dynamical mass of  
 $M_f=30$ MeV, the thin lines to $M_f=0$.   
}  
    \label{dispersion}  
  \end{center}  
\end{figure}

Alternatively, we can construct {\it approximate} analytical solutions
which have the advantage that they are more transparent.
Anti\-cipating that $\Delta_f$ will be small, we expand $E_{f;k}^2$ in
powers of $|\Delta_f|^2$ and insert this into Eq.~(\ref{GG}):
\beq
G_f(p_0,\vec p\,) = \prod_{k=3}^{6}\left(p_0^2- 
\sum_{j=0}^\infty c_{f;k}^{(j)}\, |\Delta_f|^{2j}\right)~. 
\eeq
The coefficients $c_{f;k}^{(j)}$ can be determined comparing
the r.h.s. of this expression order by order with Eq.~(\ref{G_function}).
At lowest order one recovers
the free dispersion laws for particles and antiparticles, 
$c_{f;3,5}^{(0)} = (\varepsilon_{f} - \mu_{f})^2$ and
$c_{f;4,6}^{(0)} = (\varepsilon_{f} + \mu_{f})^2$,
while the coefficients for the first non-trivial correction are obtained
as solutions of quadratic equations.  
Therefore, the particle branch and the
antiparticle branch split: 
\bea 
E_{f;3,5}^2&=& (\varepsilon_{f}-\mu_{f})^2+  
c_{f;3,5}^{(1)}\,|\Delta_{f}|^2 + \dots~, 
\nonumber 
\\  
E_{f;4,6}^2&=& (\varepsilon_{f}+\mu_{f})^2+  
c_{f;4,6}^{(1)}\,|\Delta_f|^2 + \dots~, 
\label{E_3,6} 
\eea  
where
\bea  
c_{f;3,5}^{(1)}&=&\frac{1}{2}\left[\,5-\frac{\vec p\,^2}{\varepsilon_f\mu_f}  
\pm \sqrt{\left(1-\frac{\vec p\,^2}{\varepsilon_f\mu_f}\right)^2  
+8\frac{M_f^2}{\varepsilon_f^2}}\,  
\right]~, \nonumber \\ 
c_{f;4,6}^{(1)}&=&\frac{1}{2}\left[\,5+\frac{\vec p\,^2}{\varepsilon_f\mu_f}  
\pm \sqrt{\left(1+\frac{\vec p\,^2}{\varepsilon_f\mu_f}\right)^2  
+8\frac{M_f^2}{\varepsilon_f^2}}\,  
\right]~.  \nonumber \\  
\label{coeff}  
\eea  
For illustration, we have plotted the coefficients $c^{(1)}_{f;k}$, $k=3...6$,
in Fig.~\ref{dispersion} as functions of  $|\vec p|/\mu_f$.
We have chosen a typical chemical potential $\mu_f=400$ MeV and a typical 
constituent mass $M_f=30$ MeV.  
For comparison we also show the result for $M_f=0$.
Except for small momenta, the mass effect is rather small. 

As we will see in the next section, the gap parameter $\Delta_f$
is of the order $0.1 - 10$~MeV in our model. 
We have checked numerically that Eq.~(\ref{E_3,6}),
neglecting terms of order $|\Delta_f|^4$,  
is sufficiently accurate in this case. 
We can make use of this to perform a closer examination of the 
excitation gaps in the spectrum:

Since $\Delta_f$ is small, the smallest gaps are expected in the
vicinity of the Fermi surface.
Evaluating Eqs.~(\ref{E_3,6}) and (\ref{coeff}) at the Fermi momentum,
one obtains 
\beq
    E_{f;3,5}^2({\vec p}^{\,2} = \mu_f^2 - M_f^2) 
    \simeq \left(2 \pm \sqrt{2}\,\frac{M_f}{\mu_f}\right)\,|\Delta_{f}|^2~, 
\eeq
where we have neglected higher orders in $M_f/\mu_f$. 
(The antiparticle modes $E_{f;4,6}$ are of course irrelevant for 
this discussion.)
Hence, the excitation gaps are considerably larger than 
$\Delta_{f,\rm eff}$, which is related to $E_{f;1}$, see Eq.(\ref{Deltaeff}).
In particular, none of the dispersion laws
$E_{f;3-6}$ can become gapless.

One still might worry about the fact that we have  
evaluated the energy only at the Fermi momentum. 
Performing an ``exact'' minimization, the position of the minimum becomes
shifted. However, this shift in momentum is of the order  
$|\Delta_{f}|^2$, which leads to a correction in the excitation energy
of the order $|\Delta_{f}|^4$, i.e., outside the range of validity of
our approximation. 
For instance, for $M_f = 0$, the minimum of $E_{f;5}^2$ is located at
the momentum $|\vec p| = \mu_f + |\Delta_{f}|^2/(2\mu_f)$. 
At this point one finds 
$E_{f;5}^2 = 2|\Delta_{f}|^2 - |\Delta_{f}|^4/(4\mu_f^2)$.
Of course, the last term should not be trusted.
But even if we did, we would find a gapless solution only if
$|\Delta_{f}| \geq \sqrt{8}\,\mu_f$. This is not only outside the range
of our approximations, but also far away from any realistic value
of $|\Delta_{f}|$. 

\section{Numerical results} 
\subsection{NJL model} 

As we have seen, in our model the different flavors decouple.
Therefore, the solutions of the gap equations can be discussed for
a single flavor as functions of the corresponding chemical potential
$\mu_f$. Later, we may combine the solutions for up and down quarks
(together with leptons) to obtain electrically neutral matter. 
In this article, all calculations are performed at $T = 0$.

We solve the gap equations using the dispersion relations (\ref{E_1,2}) 
for the modes $1$ and $2$ and the approximate dispersion relations
(\ref{E_3,6}) for the modes 3-6, where we neglect the terms of order
$|\Delta_f|^4$. 
The parameter sets (current quark mass $m$,  
coupling constant in the meson channel $G$,  
 and three dimensional cut-off  
$\Lambda$) are shown in Tab. \ref{parNJL}  
and have been determined by fitting the pion mass  
and the pion decay constant to their empirical values.  
They correspond to the vacuum constituent quark masses $M_f = 300$~MeV, 
$350$~MeV, and $400$~MeV, respectively. For the diquark coupling constant
we employ the color-current relation $H_v = 3G/8$.

\begin{figure}[htb]  
  \begin{center}  
    \includegraphics[width=0.82\linewidth,angle = -90]{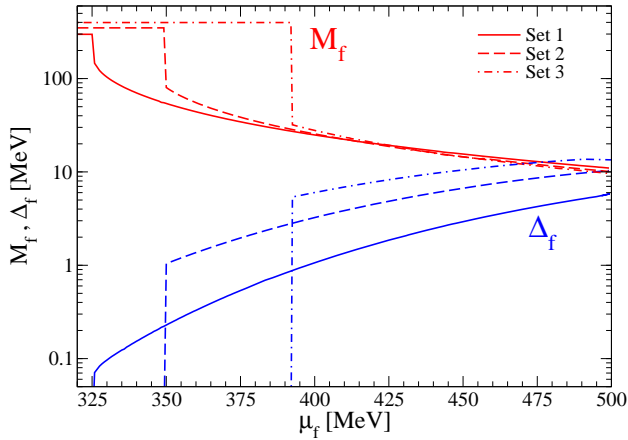}  
    \caption{Dependence of the dynamical quark mass $M_f$ and the CSL 
diquark gap $\Delta_f$  
    on the chemical potential $\mu_f$ of the quark flavor $f$ for different  
    parametrizations of the NJL model.}  
    \label{NJL2_2}  
  \end{center}  
\end{figure}  
 
\begin{table}[htb] 
\begin{center} 
\begin{tabular}{|c||c|c|c|}\hline 
Set   &$\Lambda$[MeV] &$G~\Lambda^2$ &$m$[MeV]\\ \hline 
1   &$664.3$ &$2.06$&$5.00$\\ 
2   &$613.4$ &$2.25$&$5.42$\\ 
3   &$587.9$ &$2.44$&$5.60$\\\hline 
  \end{tabular} 
\vspace{1cm} 
  \caption{Parameter sets for NJL model.} 
  \label{parNJL} 
\end{center} 
\end{table} 

In Fig. \ref{NJL2_2} we show the dynamical quark mass and the diquark 
gap as functions of $\mu_f$.
Obviously, with increasing constituent mass in vacuum, the critical
chemical potentials  $\mu_{f,{\rm crit}}$ for the chiral phase transition
also become larger. For all practical purposes, 
$\mu_{f,{\rm crit}}$ also marks the 
onset of the CSL phase.\footnote{For Set 2 and Set 3, $\mu_{f,{\rm crit}}$
is smaller than the vacuum constituent mass. Consequently, the 
density at $T = 0$ is zero for $\mu_f < \mu_{f,{\rm crit}}$, and, thus,
there are no quarks which could pair.
This is not quite the case for Set 1. Here $\mu_{f,{\rm crit}}$ is larger
than the vacuum constituent mass and, hence, there is a small interval
below $\mu_{f,{\rm crit}}$ where the density is non-zero. In this interval, 
the attractive quark-quark interaction causes a Cooper instability leading
to a non-zero CSL-condensate. However, since the density is very small,
the condensate is tiny and practically undetectable with our numerical
methods. In any case, the existence of a low-density gas of unconfined 
constituent quarks is an artifact of the model and has no physical meaning.}
Here the size of the gap    
$\Delta_f(\mu_{f,{\rm crit}})$ increases   
from Set 1 to Set 3 since the dimensionless coupling constant  
$H_v\Lambda^2$ becomes larger.     
 
The CSL gaps are strongly $\mu_f$-dependent  
functions in the considered domain   
and we easily identify two regimes:    
while  the asymptotic behavior for large density is quite similar  
for all the sets used  
($\Delta_f(\mu_{f,{\rm max}}) \simeq 10$ MeV),  
the low density region is qualitatively  
determined by the parametrization    
($\Delta_f(\mu_{f,{\rm crit}}) \simeq 0.1 - 6$~MeV depending on the set). 
In particular for Set 1 we obtain that the superconducting  
gap  has a very low value at $\mu_{f,\rm{crit}}$ 
and then rises by two orders of magnitude up to $\mu_f = 500$~MeV 
(Fig. \ref{NJL2_2}). For the other sets the corresponding  
increment is at least one order of magnitude smaller.  
  
The gap functions in the low-density region  
are  crucial  
in the derivation of equations of state for  
compact stars applications.  
Models with high $\mu_{f,{\rm crit}}$ like Set 3 might not be  
able to stabilize hybrid stars configurations \cite{Buballa:2003et}.   
 
\begin{figure}[htb]  
  \begin{center}  
    \includegraphics[width=0.82\linewidth,angle = -90]{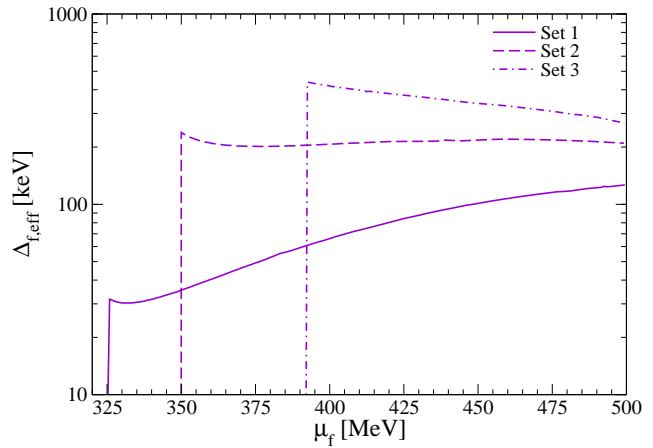}  
    \caption{Effective CSL pairing gap $\Delta_{f,{\rm eff}}$, 
             Eq.~(\ref{Deltaeff}), as function of the chemical potential
             $\mu_f$ for different parametrizations of the NJL model.} 
    \label{NJL2_1}  
  \end{center}  
\end{figure}  
 
In Fig.~\ref{NJL2_1}  
we show the results of the effective gaps $\Delta_{f,{\rm eff}}$ defined in 
(\ref{Deltaeff}) for the dispersion relations (\ref{E_1,2})  
as functions of $\mu_f$.  
As discussed in Sect.~\ref{dispers}, these are the smallest gaps in the 
spectrum and thus the relevant quantities for the compact star cooling 
phenomenology. 
As expected, the size of $\Delta_{f,{\rm eff}}$    
is up to two orders of magnitude smaller than  $\Delta_{f}$
(cf. Fig.~\ref{NJL2_2}).
We also observe that the strictly rising  
density dependence of the latter can partially be compensated by  
the drop of the dynamical quark mass $M_f$ and the increase  
of $\mu_f$. The only exception is found for Set 1,  
where  $\Delta_{f}$, as described before,  
has an early onset and rises too strongly.  
Therefore,  
we obtain that $\Delta_{f,{\rm eff}}(\mu_f)$ is  
a function which increases by  
one order of magnitude for  Set 1,  
is nearly constant  for  Set 2 and 
even decreasing for Set 3.  
 
We stress once again that  
the asymptotic behavior of the gaps at  
$\mu_{f,{\rm max}}$ is  quite similar while  
the behavior at $\mu_{f,{\rm crit}}$ 
varies  strongly  
for the different parameter sets.  
 
\subsection{Density dependent coupling} 

To obtain a reasonable description of modern cooling data,  
the pairing pattern for quark matter in hybrid star cores should 
fulfill a list of constraints, as it has been derived in a recent analysis \cite{Grigorian:2004jq}.
These were:
\begin{enumerate}
\item All quarks need to be paired (which excludes, e.g., the pure 2SC 
phase).   
\item The smallest gaps should be in the range $10 - 100$~keV 
      (which excludes, e.g., the CFL phase).
\item The smallest gaps should have a decreasing density dependence in the
relevant domain of chemical potentials
$\mu_{\rm crit} <\mu \lesssim 500$ MeV.
\end{enumerate} 
We have shown in the previous subsection that the first two conditions
are fulfilled for the CSL phase, but the third one is not. 
Therefore, to qualify the CSL phase discussed in this paper as a good
candidate for the phase structure of quark matter in  compact stars 
from the point of view of cooling phenomenology, we would like to
study here a small modification of the model which could render at
least the smallest of the 
CSL gaps a decreasing function of the density.
 
Motivated by the logarithmic momentum  dependence of the 1-loop  
running coupling constant $\alpha(\vec p\,^2)\simeq 1/\ln(\vec p\,^2
/ \Lambda^2_{{\rm QCD}})$ 
and by the fact that near the Fermi surface we have 
$|\vec p|\simeq p_F^f \simeq \mu_f$, we suggest to introduce  
a density dependence of the couplings according to 
\begin{eqnarray} 
G(\mu_f) = G R_{\Lambda_{{\rm QCD}}}(\mu_f)~,\quad
H_v(\mu_f) = H_v R_{\Lambda_{{\rm QCD}}}(\mu_f).
\label{Hvmu} 
\end{eqnarray}
Here,  the ratio 
\begin{eqnarray} 
R_{\Lambda_{{\rm QCD}}}(\mu_f)=\frac{\alpha(\mu_f)} 
{\alpha(\mu_{f,\rm crit})} 
=\frac{\ln(\mu_{f,\rm crit}/\Lambda_{{\rm QCD}})} 
{\ln(\mu_f/\Lambda_{{\rm QCD}})} ~,
\label{Rmu} 
\end{eqnarray} 
relates the running coupling constant for $\mu_f$ 
to the value at $\mu_{f,\rm crit}$. 
Of course, this formula, which has to be taken as very schematic,
cannot be used at arbitrary small chemical potentials.
In our calculations, we assume that it is valid for 
$\mu_f \geq \mu_{f,\rm crit}$ and take $G$ and $H_v$ at  
$\mu_f = \mu_{f,\rm crit}$ from Table~\ref{parNJL}.
$\Lambda_{{\rm QCD}}$ is the QCD momentum
scale which is not well known for the case $N_f=2$. We will consider
it as a free parameter to be varied in the limits   
$\Lambda_{{\rm QCD}}=210 - 300$~MeV.  
  
\begin{figure}[thb]  
  \begin{center}  
    \includegraphics[width=0.82\linewidth,angle = -90]{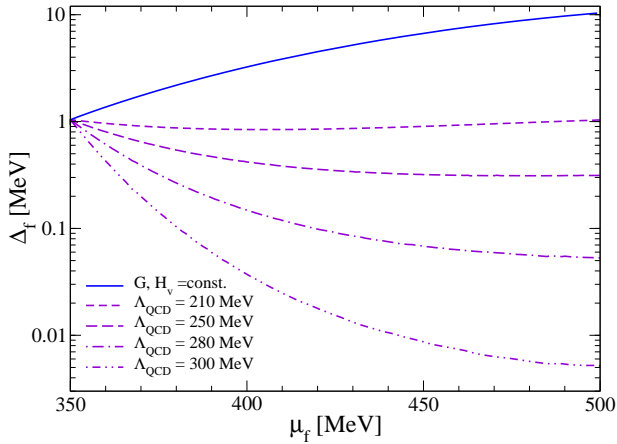}  
    \caption{ 
CSL pairing gap $\Delta_f(\mu_f)$  
for different  density dependent coupling functions 
$G(\mu_f)$, $H_v(\mu_f)$
as defined in Eq.~(\ref{Hvmu}) for Set 2  
($\mu_{f,\rm crit}= 350$ MeV).   
The solid line represents the case for constant   
$H_v = 3G/8$.  
The dashed and dash-dotted lines are for density dependent   
coupling  for  
different values of $\Lambda_{{\rm QCD}}$.} 
    \label{RunningHv}  
  \end{center}  
\end{figure}  
  
\begin{figure}[thb]  
  \begin{center}  
    \includegraphics[width=0.82\linewidth,angle = -90]{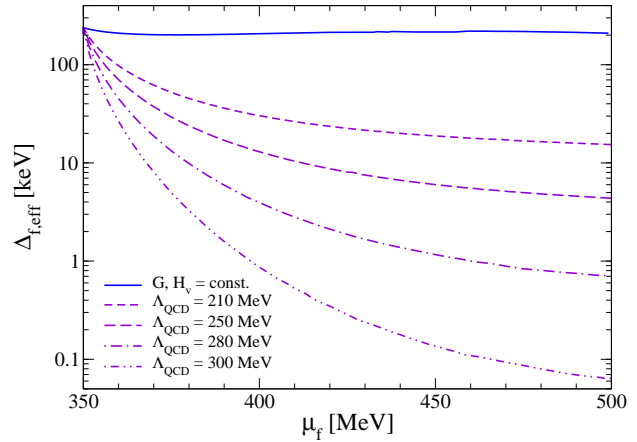}  
    \caption{Same as Fig. \ref{RunningHv} for the effective CSL gaps } 
    \label{RunningeffHv}  
  \end{center}  
\end{figure}

In Fig. \ref{RunningHv} we show the CSL gaps  
 for different $\mu_f$-dependent coupling functions  
for the Set 2 with $\mu_{\rm crit}= 350$ MeV.  
We see that the pairing gaps for $\Lambda_{{\rm QCD}}=280 - 300$ MeV  
are decreasing with $\mu_f$. 
The corresponding effective gaps shown in Fig. \ref{RunningeffHv}
change by more than two orders of
magnitude in the relevant range of chemical potentials, as it is  
required from cooling phenomenology calculations  
\cite{Grigorian:2004jq}.  
 
\section{Conclusion}  
We have employed an NJL-type model to study
a spin-1 single-flavor color superconducting phase 
which results from a color-spin locking (CSL) pairing pattern in 
two-flavor quark matter.  
This CSL phase is particularly interesting for compact star cooling  
applications since it may survive under charge neutrality constraints  
implying a sufficiently large mismatch between up- and down-quark chemical  
potentials to destroy the otherwise dominant scalar diquark condensate 
of the 2SC phase.  
 
In the specific model we used, the different flavors decouple and could 
be studied separately. 
It turned out to be important to retain non-zero quark masses, even if
they are small, because the size of the smallest gap in the spectrum
is controlled by the value of the density dependent dynamical quark mass.

Hybrid star cooling phenomenology has suggested earlier that quark 
matter in a compact star should be in a color-superconducting state 
with all quarks gapped.  
The smallest gap in the spectrum has to be in the range of several keV 
and should decrease with density.
Introducing a density dependent coupling, motivated by the running of
the QCD coupling constant, we could show that CSL effective gaps can be
brought in agreement with these requirements.
It would be interesting to see, whether this result can also be
obtained within a more microscopic approach, e.g., using Dyson-Schwinger
formalism.


Of course, before applying these results to compact stars, we must 
construct electrically neutral quark matter in beta equilibrium. 
However, since there is no pairing between different flavors,
this does not lead to major complications.
In particular, the magnitude of the CSL gaps should not change
considerably.
Therefore, since the latter nicely matches the requirements from the  
cooling phenomenology, one is encouraged to consider the CSL quark 
matter phase as a viable candidate for the superdense matter in the  
interior of compact stars.  

Another interesting question concerns the possible electromagnetic
Meissner effect \cite{bailin84,Blaschke:1999fy} in the CSL phase 
\cite{Schmitt:2003xq} in response to a strong
magnetic field typical for neutron stars. A more detailed discussion of
this issue is, however, beyond the scope of the present work.
 
\section*{Acknowledgment} 
We are grateful to  H. Grigorian,  D. K. Hong, D. Rischke, T. Sch\"afer, 
A. Schmitt, I. Shovkovy, D. Voskresensky and Q. Wang for their 
comments and interest in our work. 
We acknowledge discussions during the meetings of the Virtual Institute  
VH-VI-041 of the Helmholtz Association on 
{\it Dense hadronic matter and QCD 
phase transition} and the Helmholtz International Summer School  
{\it Hot points in Astrophysics}, Dubna, August 2-12, 2004.  
D.N.A. acknowledges support from Landesgraduiertenf\"orderung 
Mecklenburg-Vorpommern, the work of V.Y. was supported in part by DFG under grant No. RUS 117/37/03, by the Dynasty Foundation and by RFBR grant No. 05-02-16699.
D.B. and M.B. thank for partial support of the Department of Energy 
during the program INT-04-1 on 
{\it QCD and Dense Matter: From Lattices to Stars}  at the 
University of Washington, where this project has been initiated.


\begin{thebibliography}{99}  
  
\bibitem{Rapp:1997zu}
  R.~Rapp, T.~Sch\"afer, E.~V.~Shuryak and M.~Velkovsky,
  Phys.\ Rev.\ Lett.\  {\bf 81}, 53 (1998)

\bibitem{Alford:1997zt}
  M.~G.~Alford, K.~Rajagopal and F.~Wilczek,
  Phys.\ Lett.\ B {\bf 422}, 247 (1998)
  
\bibitem{Blaschke:1998md} 
  D.~Blaschke and C.~D.~Roberts, 
  Nucl.\ Phys.\ A {\bf 642} (1998) 197. 

\bibitem{barrois77}  
B.C. Barrois, Nucl. Phys. {\bf B129} (1977) 390.  

\bibitem{bailin84}  
D. Bailin and A. Love, Phys. Rep. {\bf 107} (1984) 325.  
 
\bibitem{Alford:2001dt}  
M.~G.~Alford,  
Ann.\ Rev.\ Nucl.\ Part.\ Sci.\  {\bf 51} (2001) 131.  
 
\bibitem{Buballa:2003qv} 
  M.~Buballa, 
  Phys. Rep. {\bf 407} (2005) 205.
  
\bibitem{Schmitt:2004et}
  A.~Schmitt,
  Phys.\ Rev.\ D {\bf 71} (2005) 054016.

\bibitem{Ruster:2005jc}
  S.~B.~R\"uster, V.~Werth, M.~Buballa, I.~A.~Shovkovy and D.~H.~Rischke,
  arXiv:hep-ph/0503184.

\bibitem{Blaschke:2005uj}
  D.~Blaschke, S.~Fredriksson, H.~Grigorian, A.~M.~\"Oztas and F.~Sandin,
  arXiv:hep-ph/0503194.
 
\bibitem{Neumann:2002jm}  
F.~Neumann, M.~Buballa and M.~Oertel,  
Nucl.\ Phys.\ A {\bf 714}, 481 (2003).  
 
\bibitem{Oertel:2002pj} 
  M.~Oertel and M.~Buballa, 
  arXiv:hep-ph/0202098. 

\bibitem{Alford:2002kj}  
M.~Alford and K.~Rajagopal,  
JHEP {\bf 0206} (2002) 031.  




\bibitem{Gocke:2001ri} 
  C.~Gocke, D.~Blaschke, A.~Khalatyan and H.~Grigorian, 
  arXiv:hep-ph/0104183. 
 
\bibitem{Grigorian:2003vi} 
  H.~Grigorian, D.~Blaschke and D.~N.~Aguilera, 
  Phys.\ Rev.\ C {\bf 69}, 065802 (2004). 
 
\bibitem{Shovkovy:2003ce} 
  I.~Shovkovy, M.~Hanauske and M.~Huang, 
  Phys.\ Rev.\ D {\bf 67}, 103004 (2003). 
 
\bibitem{Aguilera:2004ag} 
  D.~N.~Aguilera, D.~Blaschke and H.~Grigorian, 
  Nucl. Phys. A (2005) in press, arXiv:hep-ph/0412266. 


 
\bibitem{Blaschke:1999qx} 
  D.~Blaschke, T.~Kl\"ahn and D.~N.~Voskresensky, 
  Astrophys.\ J.\  {\bf 533} (2000) 406. 
 
\bibitem{Page:2000wt} 
  D.~Page, M.~Prakash, J.~M.~Lattimer and A.~Steiner, 
  Phys.\ Rev.\ Lett.\  {\bf 85} (2000) 2048. 
 
\bibitem{Blaschke:2000dy} 
  D.~Blaschke, H.~Grigorian and D.~N.~Voskresensky, 
  Astron.\ Astrophys.\  {\bf 368} (2001) 561. 
 
\bibitem{Grigorian:2004jq} 
  H.~Grigorian, D.~Blaschke and D.~Voskresensky, 
  Phys. Rev. C (2005) in press;   
  arXiv:astro-ph/0411619. 
 
\bibitem{Schafer:2000tw}  
T.~Sch\"afer,  
Phys.\ Rev.\ D {\bf 62} (2000) 094007.  
  
\bibitem{Alford:2002rz}  
M.~G.~Alford, J.~A.~Bowers, J.~M.~Cheyne and G.~A.~Cowan,  
Phys.\ Rev.\ D {\bf 67} (2003) 054018.  
  
\bibitem{Schmitt:2005ee} 
  A.~Schmitt, I.~A.~Shovkovy and Q.~Wang, 
  arXiv:hep-ph/0502166. 
 
\bibitem{Buballa:2003et}
  M.~Buballa, F.~Neumann, M.~Oertel and I.~Shovkovy,
  Phys.\ Lett.\ B {\bf 595}, 36 (2004).

\bibitem{Blaschke:1999fy}
  D.~Blaschke, D.~M.~Sedrakian and K.~M.~Shahabasian,
  Astron.\ Astrophys.\  {\bf 350} (1999) L47.

\bibitem{Schmitt:2003xq}
  A.~Schmitt, Q.~Wang and D.~H.~Rischke,
  Phys.\ Rev.\ Lett.\  {\bf 91} (2003) 242301.


\end{thebibliography}
\end{document}